\documentclass[twocolumn,english,prb,showpacs,preprintnumbers,amsmath,amssymb]{revtex4}
\usepackage{graphicx}% Include figure files
\usepackage{dcolumn}% Align table columns on decimal point
\usepackage{bm}% bold mat
\usepackage{amsbsy}

\makeatletter
\@ifundefined{textcolor}{}
{%
 \definecolor{BLACK}{gray}{0}
 \definecolor{WHITE}{gray}{1}
 \definecolor{RED}{rgb}{1,0,0}
 \definecolor{GREEN}{rgb}{0,1,0}
 \definecolor{BLUE}{rgb}{0,0,1}
 \definecolor{CYAN}{cmyk}{1,0,0,0}
 \definecolor{MAGENTA}{cmyk}{0,1,0,0}
 \definecolor{YELLOW}{cmyk}{0,0,1,0}
 }

\usepackage{epsfig}

\usepackage{dcolumn}% Align table columns on decimal point
\usepackage{bm}% bold math
\usepackage{wasysym}

\usepackage{marvosym}

\@ifundefined{definecolor}
 {\usepackage{color}}{}

\usepackage{hyperref}

\makeatletter

\usepackage{ulem}
\makeatother

\makeatother

\usepackage{babel}

\makeatother

\usepackage{babel}

\begin{document}

\title{The role of pressure on the magnetism of bilayer graphene}

\author{ Eduardo V. Castro,$^{1,2}$ Mar\'ia P. L\'opez-Sancho,$^{1}$, and Mar\'ia A.
H. Vozmediano$^{1}$}

\affiliation{$^{1}$Instituto de Ciencia de Materiales de Madrid, CSIC, Cantoblanco,
E-28049 Madrid, Spain}

\affiliation{$^{2}$Centro de F\'isica do Porto, Rua do Campo Alegre 687, P-4169-007
Porto, Portugal}

\begin{abstract}

We study the effect of pressure on the localized magnetic moments induced by vacancies 
in bilayer graphene in the presence of topological defects breaking the
bipartite nature of the lattice. 
By using a mean-field Hubbard model we address the two inequivalent types of vacancies
that appear in the Bernal stacking bilayer graphene.
We find that by  applying pressure in 
the direction perpendicular to the layers the critical value of the Hubbard interaction needed to polarize 
the system decreases. The effect is particularly enhanced for one type of vacancies, and admits straightforward generalization to multilayer graphene in
Bernal stacking and graphite. The present results clearly demonstrate that the
 magnetic behavior of multilayer
graphene can  be affected  by mechanical transverse deformation.

\end{abstract}
\maketitle

\section{Introduction}

The magnetic properties of graphene and its multilayer compounds remain one of the most 
interesting topics in the system that still awaits experimental confirmation. 
The improved experimental capabilities to produce and manipulate large samples 
has given rise to a renewed interest on the issue.\cite{Kimetal09} 
It is known that vacancies, edges and other defects that lead to dangling bonds 
in the graphene system induce localized magnetic moments that might give 
rise to interesting applications. The theoretical paradigm around the magnetism 
in graphene is the Lieb theorem\cite{L89} that fixes the spin of the ground state 
of the bipartite system to be half the number of the unpaired atoms in the lattice. 
Although the theorem is demonstrated only for the Hubbard model 
in bipartite lattices, the result has proven 
to be very robust and to hold in more general calculations based on ab initio or density 
functional methods\cite{MLFN04,PFRB08}. While the interactions can be extended, 
it has been recently proven that the ground state magnetization is very sensitive to 
the presence of local topological defects such as five or seven rings breaking the bipartite 
character of the lattice.\cite{Cetal08,SSAR10}
In particular it has been shown that, when one of the vacancies inducing magnetic moments 
is reconstructed to form a pentagon, 
the critical value of the Hubbard 
interaction needed to reach a finite polarization 
increases significantly.\cite{LJV09}

Bilayer graphene (BLG) is even more interesting than single layer graphene
(SLG) under many points of view,\cite{NGPNG09}
in particular for the magnetic properties. 
In the Bernal-stacking, BLG can support two types 
of vacancies giving rise to unpaired atoms: 
these produced by removing a site having
a neighbor in the adjacent layer are named  
$\beta$, and these coming from sites that are not connected to the other layer 
called $\alpha$ vacancies. In a bipartite lattice, unpaired atoms give rise
to zero energy states and the physics underlying the magnetic properties of the system
is that of the electronic interactions in the manifold
of zero energy states. It has been recently  shown \cite{CLV09,CLV10}  
that the significant differences between the wave functions of the zero modes 
associated to the $\alpha$ and
$\beta$ vacancies in BLG give rise to different physical behaviors. 
In particular when the system is gapped by an external gate, the vacancies  
of type $\alpha$ generate fully localized states inside the gap.

Recent experimental progress in production and 
manipulation of graphene samples have broadened the possibilities 
of tailoring the properties of SLG and BLG. Pressure is known to play
an important role on the stability of the gate induced gap in BLG \cite{GGC08}
and on the impurity states \cite{DBZ08}.

In this work we analyze the effect of pressure on the behavior of the localized magnetic 
moments coming from the two types of vacancies in the presence of a topological defect. 
We show that  
the critical interaction value $U_c$ needed to polarize the system decreases for increasing pressure. 
The effect is particularly enhanced for vacancies of type $\alpha$. We clarify  the physical mechanism 
leading to this behavior and propose pressure as a way to improve the magnetism of the samples.

We work with the tight-binding (TB) model, including a Hubbard interaction. 
The results are obtained by a self-consistent computation within the unrestricted
Hartree-Fock approximation. A further analytical analysis using
degenerate first order perturbation theory explains the results in terms of
the different energy  lifting that occurs in the subspace of the zero modes 
under the perturbations originated by the pentagonal hopping.
We argue that the physical result will remain when going beyond the simple 
calculation performed in this work and extend the results to other graphene  
multilayers.

\begin{figure}[t]
\begin{centering}
\includegraphics[width=0.99\columnwidth]{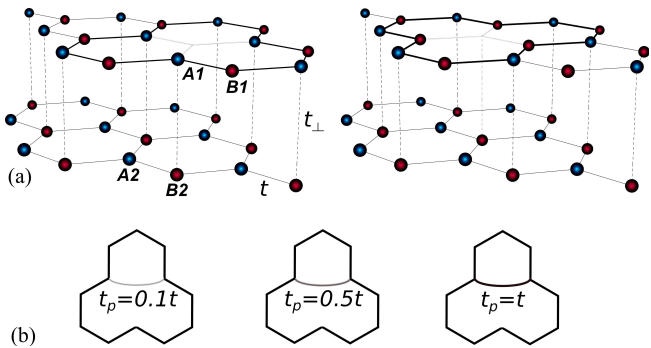} 
\par\end{centering}

\caption{\label{fig:lattice}(Color online) (a)~Bilayer lattice structure
and main tight-binding parameters. Left: $\alpha$ vacancy. Right: $\beta$ 
vacancy. (b)~Reconstructed vacancies modeled by a pentagonal link.}

\end{figure}

\section{Model}

The lattice structure of AB Bernal-stacking BLG
is schematically represented in Fig. ~\ref{fig:lattice}(a). 
The atoms of the A sublattice in the 
top layer are connected by $t_\perp$ to these of the
B sublattice of the bottom layer. We will consider the minimal
model for the BLG with only intralayer  $t$ and inter layer $t_\perp$
couplings.
The estimated values of these  parameters in the system are
$t\approx 3\,\mbox{eV}$, and
$t_\perp\approx0.3-0.4\,\mbox{eV}\sim t/10$ \cite{NGPNG09}.

The two different  types
of vacancies are also shown in Fig. ~\ref{fig:lattice} (a).
The $\alpha$ vacancy where the removed atom in the top layer ($B_1$ in the figure)
is not connected to the second layer is represented in the left hand side. The
right hand side represents a  $\beta$ vacancy ($A_1$ in the figure). Fig. ~\ref{fig:lattice}
(b) shows the reconstructed pentagonal link named $t_p$ in this work.

The non-interacting TB Hamiltonian for the $\pi$-electrons of the bilayer system
is:\cite{McClure57,SW58,MF06}
\begin{equation}
H_{TB}=\sum_{i=1}^{2}H_{i}-t_\perp \sum_{\mathbf{R},\sigma}
\big[a_{1\sigma}^{\dagger}(\mathbf{R})b_{2\sigma}(\mathbf{R})+
\mbox{h.c.}\big],
\label{eq:Hbilayer}
\end{equation}
where $H_i$ corresponds to the SLG Hamiltonian:
%
%\begin{widetext}
\begin{multline}
H_{i}=\\
-t\sum_{\mathbf{R},\sigma} a_{i\sigma}^{\dagger}(\mathbf{R})
\big[
b_{i\sigma}(\mathbf{R})+
b_{i\sigma}(\mathbf{R}-\mathbf{a}_{1})+
b_{i\sigma}(\mathbf{R}-\mathbf{a}_{2})\big] +\mbox{h.c.}\,,
\label{eq:Hslg}
\end{multline}
%\end{widetext}
%
and $a_{i\sigma}(\mathbf{R})$ {[}$b_{i\sigma}(\mathbf{R})$] are
the annihilation operators for electrons at position $\mathbf{R}$
in the sublattice $A_i$ ($B_i$) of the layer $i$ ($i=1,2$) with spin polarization  $\sigma$.
The basis vectors can be chosen as
$\mathbf{a}_{1}=a\,\rm{\hat e}_{x}$ and
$\mathbf{a}_{2}=a(\rm{\hat e}_{x}-\sqrt{3}\,\rm{\hat e}_{y})/2$,
with $a=0.246\,\rm{nm}$ being the lattice spacing.

The lattice of 
BLG is a bipartite lattice
and the TB minimal model  has electron-hole symmetry.
Vacancies are modeled by suppressing  the corresponding lattice site.
No reconstruction or relaxation 
will be included in the remaining structure which maintains its original geometry. 
This approximation does not affect the zero-modes we are interested in \cite{CJKK08}.
We will consider the simplest topological defect  modeled by first producing a vacancy and then
adding a pentagonal link connecting two of the closest atoms to the vacancy
as shown in Fig.~\ref{fig:lattice}(b). Since the two atoms belong to the same
sublattice this defect breaks the bipartite nature of the lattice.

The magnetic behavior of  BLG in the presence of vacancies, edges,
and other defects has been investigated using the Hubbard model in 
\onlinecite{CPSjoam07,CLV09,CPS+07,SCS+08}.
The interacting TB Hamiltonian is $H=H_{TB}+H_{U}$, 
with
\begin{equation}
H_{U}=U\sum_{\bm R,\iota}[n_{a\iota\uparrow}(\bm R)n_{a\iota\downarrow}
(\bm R)+n_{b\iota\uparrow}(\bm R)n_{b\iota\downarrow}(\bm R)]\,,
\label{eq:Hubb}
\end{equation}
where $n_{x\iota\sigma}(\bm R)=x_{\iota\sigma}^{\dag}(\bm
R)x_{\iota\sigma}(\bm R)$, with $x=a,b$, $\iota=1,2$ and
$\sigma=\uparrow,\downarrow$. 

We use finite clusters with periodic
boundary conditions at half-filling (one electron per atom).
The Hamiltonian is solved in the Hartree-Fock approximation, 
and
the mean-field spin density at each lattice site is obtained
self-consistently. It is known that a finite  
staggered magnetization  appears in the honeycomb lattice
above a critical value of the on-site Coulomb interaction $U_{c}\approx 2.2t$
\cite{Fetal96,PAB04}. 
We will keep the values of  $U$ below this value to make sure that 
the physics explored is due to the magnetic moments associated to the
defects on the lattice. Evidence of the presence of localized
magnetic moments around isolated vacancies has been reported recently in 
scanning tunneling microscopy experiments 
 on a graphite surface \cite{UBGG10}.
 
\section{Vacancy-induced zero-energy states}
\label{Sect_wavefuncs}

In the honeycomb lattice, a vacancy gives rise to a quasilocalized state
with a continuum limit wave function
\begin{equation}
\Psi(x,y)\approx
\frac{e^{i\mathbf{K}.\mathbf{r}}}{x+iy}+\frac{e^{i\mathbf{K}'.\mathbf{r}}}{x-iy},
\label{eq:PsiVac1L}
\end{equation}
where $\mathbf{K}$ and $\mathbf{K}'$ are the reciprocal space vectors
of the two inequivalent corners of the first Brillouin zone, and $(x,y)$
are distances from the vacancy position.\cite{PGS+06}

An analytic expression for the vacancy-induced states in the continuum model of BLG was 
obtained recently in 
Ref.~\onlinecite{CLV10} following the procedure outlined in 
\onlinecite{CPetal08,CLV10} for the monolayer case. 
By cutting the lattice into left and right regions with respect to the vacancy 
position a zigzag edge to the left and a Klein edge to the right appear. 
The wave function is obtained by matching surface state solutions 
at the zigzag edge with that localized at the Klein edge.
Depending on the type of vacancy,
two solutions are obtained.
The $\beta$ vacancy produces a zero-energy state quasilocalized
 on atoms of the opposite sublattice 
in the same layer of the vacancy, decaying as $1/r$ away from the vacancy. 
Assuming the vacancy to be located in layer~1, the 
continuum limit wave function can be written as
\begin{eqnarray}
\Upsilon_1(x,y) & \approx & \Psi(x,y) \nonumber \\
\Upsilon_2(x,y) & \approx & 0\,,
\label{eq:betawf}
\end{eqnarray}
where $\Upsilon_i$ specifies the wave function component on layer~$i$, and
$\Psi(x,y)$  is the quasilocalized state given in Eq.~\eqref{eq:PsiVac1L}.
The zero-energy states induced by a vacancy of  $\alpha$ type has
the wave function
\begin{eqnarray}
\Upsilon_1(x,y) & \approx & \Psi(x,y) \nonumber \\
\Upsilon_2(x,y) & \approx & 
\frac{t_\perp}{t}e^{-i2\theta}e^{i\mathbf{K}.\mathbf{r}}+
\frac{t_\perp}{t}e^{i2\theta}e^{i\mathbf{K}'.\mathbf{r}}\,,
\label{eq:alphawf}
\end{eqnarray}
 where $\theta = \arctan(y/x)$.
This is a delocalized state, with the peculiarity of being quasilocalized
in one layer (where the vacancy sits) and delocalized in the other \cite{CLV10}.
We notice here that the wave function of the $\beta$ vacancy is insensitive to
$t_\perp$ while this parameter enters explicitly in the wave function of the 
$\alpha$ vacancy, a fact that will be important in the forthcoming analysis.

\section{Numerical results}

\subsection{Magnetism and the pentagonal links}
Since the lattice of the BLG model used in this work is
bipartite the magnetic behaviour of the system follows the Lieb's theorem
as in SLG. Therefore in the presence of vacancies, for a repulsive value of the Hubbard
interaction $U$ the ground state of the system at half filling has a total 
spin equal to half the number of unbalanced atoms $S=(N_A-N_B)/2$.
We have analysed the magnetic behaviour of  the two different types of 
vacancies that occur  in the BLG in the presence of a pentagonal link.    

We observed that both types of vacancies have the same behaviour
and it coincides with what happens in SLG \cite{LJV09}.
We have considered two  vacancies on the same layer and  reconstructed 
one of them by forming a pentagonal link $t_p$
as pictorially shown in Fig.~\ref{fig:lattice}(b). 
In this situation a finite
critical value of the on-site Coulomb interaction is needed to
reach the ground state polarization predicted by Lieb's theorem. 
For values of $U$ below the $U_c$ the total spin of the ground state is zero.
By varying the pentagonal hopping integral 
from $0.1t$ to $t$ we have verified that the critical value of $U$ 
increases monotonically with $t_p$ for both types of vacancies.

\begin{figure}[t]
\begin{centering}
\includegraphics[width=0.90\columnwidth]{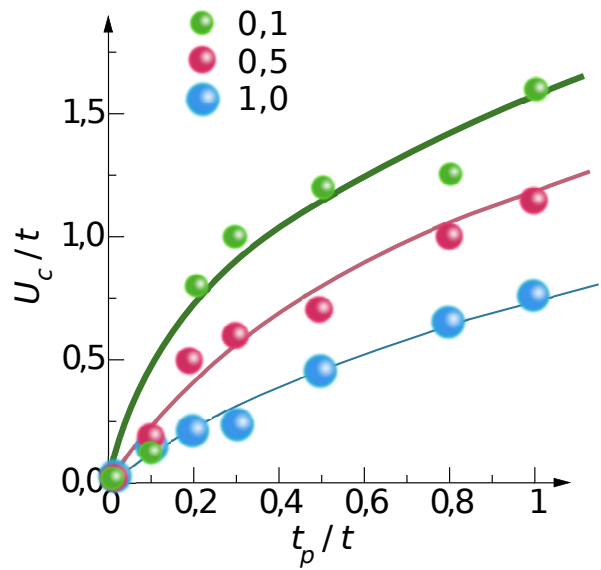} 
\par\end{centering}
\caption{\label{fig:results} (Color online) Critical  value of the Hubbard parameter
$U$ separating zero
from finite polarization regions as a function of the strength
of the pentagonal link$t_p$, for different values of the perpendicular 
hopping $t_\perp$ in units of $t$. The dots are computed values; lines are a guide to the eye.}
\end{figure}

\subsection{The role of pressure}

We have explored the influence of pressure on the magnetic properties
of BLG. The effect of increasing pressure is modelled by 
increasing the hopping between atoms of different layers $t_\perp$.
The most common value for $t_\perp$  considered in BLG is $t_\perp \approx 0.1t$
although recent works considering 
different types of stacking in few layers graphene have produced 
values  up to $t_\perp \sim 0.502$  eV
with $t = 3.16$ eV \cite{ZSMMD10}.

In order to exemplify the behaviour found in this work and 
for simplicity we will study the magnetic polarization of the system in the 
situation described before: Two vacancies of the same type,
one of them reconstructed forming a pentagonal ring. 
We have seen that the effect of pressure is very different 
for the two different types of vacancies of BLG.

We have varied the perpendicular hopping  in the range  $t_\perp= 0.1t$ to
$t_\perp=t$ and for each value of $t_\perp$ we increased the
pentagonal hopping $t_p$ from $0.1t$ to $t$. We found that 
the magnetic behavior of the
system is independent on  $t_\perp$. The value of the $U_c$
increases with $t_p$ but does not depend on  $t_\perp$. 
Pressure has no effects on the spin polarization of 
the system in this case. This can be understood by 
observing that, since
the wave functions induced
by $\beta$ vacancies have amplitude only on one layer 
the behavior is just that of the SLG.
This behavior will change if next nearest
interlayer hoppings $\gamma_3$ and $\gamma_4$ are taken into account,
with $\beta$ vacancies becoming more similar to $\alpha$ vacancies.

Considering  $\alpha$  vacancies we observed that for a fixed value of the pentagonal hopping $t_p$,
$U_c$  decreases as $t_\perp$ increases. This indicates that pressure could help
the polarization of the BLG ground state.
The critical values of the 
on-site Coulomb interaction
are represented against the pentagonal hopping
in  Fig. \ref{fig:results}. Each curve
corresponds to a different value of $t_\perp$. While the strength of the 
pentagonal hopping conspires against the polarization, the perpendicular
hopping favors the appearance of magnetic moments.
In the next section Sec.~\ref{sec:pt} we
explain this behavior doing an analysis with degenerate perturbation theory.

We have also considered  case of two vacancies in
different layers. To obtain a non zero magnetization according to
the Lieb's theorem the two vacancies must belong to the same sublattice, A or B. 
Therefore in this case to get a magnetic ground state
both types of vacancies $\alpha$ and $\beta$ must be  present.
We observe that $t_\perp$ has no effect on the critical value of 
the Hubbard interaction needed to obtain the spin polarization
of the ground state
when the pentagonal link is attached to a vacancy of $\beta$ type.
On the contrary, when the pentagonal ring
is formed by reconstructing the $\alpha$ vacancy $U_c$ decreases as
$t_\perp$ increases as it occurs in the case of having  both vacancies
of $\alpha$ type.
Therefore the increase of $t_\perp$ affects the magnetic 
behavior when a  pentagonal hopping exists. Otherwise the
Lieb's theorem holds for any value of $t_\perp$. 

%{\bf Do we give any details on the numerical work? Number of lattice sites, dependence on the cluster's size, comments on what happens when the pentagon is far away, etc.}

\section{Perturbative analysis}
\label{sec:pt}

In the following we explain the results obtained in the previous section by
studying how the zero-energy modes introduced in Sect \ref{Sect_wavefuncs}
are  affected by the pentagonal link giving rise to the topological defect shown 
in Fig.~\ref{fig:lattice}(b). We use first order degenerate perturbation
theory. The discussion is put on general grounds having in mind application 
to multilayer graphene and graphite.

\subsection{Definitions}

Let $H$ be the non-interacting TB model Hamiltonian
for some graphitic system preserving the bipartite nature of the
lattice, as is the case of Eq~\eqref{eq:Hbilayer}. 
Let as assume the system to hold $n$ vacancies belonging
to the same sublattice, with $n$ induced zero energy modes 
$\left|\psi_{v}^{1}\right\rangle \dots\left|\psi_{v}^{n}\right\rangle $,
such that
\begin{equation}
H\left|\psi_{v}^{i}\right\rangle =0.\label{eq:zm}
\end{equation}
Explicitly, using localized atomic orbitals as TB basis,
$\left\{ \left|1\right\rangle ,\dots,\left|N\right\rangle \right\} $,
where $N$ is the number of lattice sites in the system, we can write
the zero energy modes as
\begin{equation}
\left|\psi_{v}^{i}\right\rangle =\sum_{j}a_{j}\left|j\right\rangle.
\label{eq:zmatomic}
\end{equation}
Without loss of generality, we assume the vacancies to belong to the
A~sublattice, which allows us to write
\begin{equation}
\left|\psi_{v}^{i}\right\rangle =\sum_{j\in B}a_{j}\left|j\right\rangle.
\label{eq:zmatomicB}
\end{equation}
In the case of SLG, for example, the continuum limit would
give $\langle \bm r \left|\psi_{v}^{i}\right\rangle \approx \Psi(x,y)$, with
$\Psi(x,y)$ as given in Eq.~\eqref{eq:PsiVac1L}.

We want to perturb the system with a term that disrupts the
bipartite nature of the lattice. The simplest such term is just an
extra local hopping connecting the same sublattice,
\begin{equation}
H_\gamma =t_{p}c_{\gamma}^{\dagger}c_{\gamma+\bm{\delta}},
\label{eq:Hp}
\end{equation}
where $c_{\gamma}^{\dagger}$ creates an electron
in the localized atomic orbital $\left|\gamma\right\rangle $,
and $\bm{\delta}$ is a vector connecting next nearest neighbors.
The local perturbation defined by Eq.~\eqref{eq:Hp} provides a
simple parametrization of the pentagonal link shown 
in Fig.~\ref{fig:lattice}(b).

\subsection{First order degenerate perturbation theory}

We want to know what happens to zero energy modes once we add the
perturbation defined by Eq.~\eqref{eq:Hp} to the system.
Here we use a perturbative analysis, and apply degenerate first
order perturbation theory within the zero energy mode sector. This
is justified, strictly speaking, if the perturbing parameter $t_{p}$
is much smaller than the HOMO-LUMO gap.

The new energies for $\left|\psi_{v}^{i}\right\rangle $ within first
order perturbation theory are given by the eigenvectors of the matrix
\begin{equation}
\mathcal{T}=\left(\begin{array}{ccc}
T_{11} & T_{12} & \dots\\
T_{21} & T_{22} & \dots\\
\vdots & \vdots & \ddots\end{array}\right),
\label{eq:1storder}
\end{equation}
where 
\begin{equation}
T_{ij}=\left\langle \psi_{v}^{i}\right|H_\gamma\left|\psi_{v}^{j}\right\rangle 
=\begin{cases}
0 & \gamma \in A\\
t_{p}a_{\gamma}^{*}a_{\gamma+\bm{\delta}} & \gamma \in B\end{cases},
\label{eq:elements}
\end{equation}
as we have assumed the vacancies to belong to the A~sublattice.

Owing to the quasilocalized nature of vacancy induced zero modes (in
multilayer systems we assume $H_\gamma$ to act on the layer where
the zero mode has its quasilocalized component) we may consider the
limiting case where vacancies are sufficiently far apart. In this particular
case zero modes will be almost unaffected if $t_{p}$ connects two
sites that are also sufficiently far apart from any vacancy. Once these
two sites approach a given vacancy we will see the energy of the associated
zero mode go up as
\begin{equation}
E\approx t_{p}a_{\gamma}^{*}a_{\gamma+\bm{\delta}},
\label{eq:Eshift}
\end{equation}
where $a_{\gamma}$ and $a_{\gamma+\bm{\delta}'}$ are the real amplitudes
of that particular zero mode at the perturbed sites. The other zero
modes being almost unaffected. This is certainly a good description
for the pentagonal link shown in Fig.~\ref{fig:lattice}(b), when
$t_p$ connects two of the closest sites to a given vacancy.

It was shown in Ref.~\onlinecite{LJV09} that in SLG the pentagonal
link induces a finite Hubbard interaction $U_c$ to polarize the
system. Within the present perturbative analysis this can be understood
as a consequence of the energy shift of the zero mode affected by the
pentagonal link: the effect of the Hubbard term has to overcome the
energy scale set by the shift in Eq.~\eqref{eq:Eshift}.

Multilayer graphene, and in particular BLG, are of especial
interest. There we can tune $a_{\gamma}$ by applying pressure, since
by increasing hopping between layers the amplitude of the zero mode
over the layer where the vacancy resides decreases. This is apparent
in Eq.~\eqref{eq:alphawf} for zero mode induced by the $\alpha$
type vacancy and explains the numerical results shown in Fig. \ref{fig:results}.

\section{Summary and discussion}
The possibility of tuning
magnetic behavior by lattice deformation in graphitic materials is nowadays being considered.
Tunable magnetism by mechanical control in graphene
is a hot topic due to the broad field of applications of organic magnets
and, from the conceptual point of view a challenge.
In this work we
have shown  that  pressure applied in the perpendicular direction
to the planes can override the negative effect of topological defects breaking the
sublattice symmetry in multilayer graphene. 

Vacancies are the principal type of defects produced in graphitic materials by 
ion bombardment \cite{TDetal08,TB09} and play a substantial role in material properties. Topological defects
breaking the sublattice symmetry are also energetically favorable and are being considered 
in defect engineered devices \cite{Hetal04,Metal08,YL10}.

We have seen that the two different types of vacancies that can form in Bernal stacked 
multilayer graphene have different magnetic behaviors under pressure. Both of them
are affected by topological defects in the sense that a larger value of the Hubbard interaction is needed to polarize the ground state. We have seen that the critical $U$ decreases by applying pressure to the sample in the case of having a majority of $\alpha$ vacancies. The $\beta$ vacancies are not affected by pressure.  
These different behaviors are due to 
the differences in the wave functions of the zero-energy states induced by vacancies of the two types.
 
%{\bf Say something on the behavior as a distance pentagon-vacancies etc. On the trilayer. On seral vacancies etc.}

\begin{acknowledgments}
We thank Fernando de Juan for useful discussions. Support from
MEC (Spain) through grants FIS2008-00124, PIB2010BZ-00512  is acknowledged.
\end{acknowledgments}
\bibliography{Presion}
\end{document}